\documentstyle[12pt]{article}
\newcommand{\be}{\begin{equation}}
\newcommand{\ee}{\end{equation}}
\newcommand{\bea}{\begin{eqnarray}}
\newcommand{\eea}{\end{eqnarray}}
\newcommand{\ba}{\begin{array}}
\newcommand{\ea}{\end{array}}

\newcommand{\bc}{\begin{center}}
\newcommand{\ec}{\end{center}}
\newcommand{\ben}{\begin{enumerate}}
\newcommand{\een}{\end{enumerate}}
\newcommand{\bfi}{\begin{figure}}
\newcommand{\efi}{\end{figure}}

\newcommand{\bq}{\begin{quote}}
\newcommand{\eq}{\end{quote}}
\newcommand{\bqu}{\begin{quotation}}
\newcommand{\equ}{\end{quotation}}
\newenvironment{emphit}{\begin{itemize}}{\end{itemize}}
\newcommand{\bemp}{\begin{emphit}}
\newcommand{\eemp}{\end{emphit}}

\newcommand{\bt}{\begin{tabular}}
\newcommand{\et}{\end{tabular}}

\date{ \ \ }
\begin{document}
\begin{center}
{\bf {\Large PRESENTATION THEOREMS FOR CODED CHARACTER SETS}}\\
\ \\
{\bf{\large Dele Oluwade}}\\
P.O.Box 20253,\\
University of Ibadan,\\
Ibadan,OYO 200005,\\
Nigeria\\
e--mail: oluwade@sigmaxi.net
\end{center}
\newpage
{\bf Abstract.}

 The notion of 'presentation', as used in combinatorial group theory,
is applied to coded character sets(CCSs) $-$ sets which facilitate
the interchange of messages in a digital computer network(DCN) .
By grouping each element of the set into two portions and using
the idea of group presentation( whereby a group is specified by
its set of generators and its set of relators) , the presentation
of a CCS is described. This is illustrated using the Extended
Binary Coded Decimal Interchange Code(EBCDIC) which is one of the
most popular CCSs in DCNs.

\newpage
{\bf 1. Introduction.}

Most of the data which are presented to a digital computer
system(DCS) are in the form of records and these records are
usually entered as alphanumeric characters. For every number which
is manipulated by a DCS, ten alphabetic characters are processed
[13]. Mathematically, a coded character set(CCS) arises as a
result of a mapping between the set of binary digits(bits) and the
set of characters ; the emphasis in this work is on the set of
bits(codes) . Thus, the set of bits of a CCS, and in effect a CCS,
is structurally a set of sequences. The ordering in the set is
according to the collating sequence( i.e  the natural sequence of
appearance in the set). Coded character sets(CCSs) are important
in a DCS because they provide a means of representing alphanumeric
characters(i.e numerals, letters, punctuation characters and
control codes) as fixed sequences of zeros and ones. The sets are
normally utilized in the sixth layer(i.e presentation layer) of
the seven open system interconnection(OSI) layers of computer
network [34]. Two popular examples of such sets are the $7-bit$
American Standard Code for Information Interchange(ASCII)
consisting of 128 characters and the Extended Binary Coded Decimal
Interchange Code(EBCDIC) consisting of 256 characters. Although a
CCS is a code(i.e it is a code mapped unto a set of characters),
it is not an error detection or correction(EDC) code and also, it
is not always a group code . {A well-known result is that a code
is linear iff it is a group [4] }. A lot of the literature has
been devoted to studies of the mathematical properties of linear
and nonlinear codes and of the construction of EDC codes e.g see
[1, 10, 11, 19]. In particular, some properties of group codes are
discussed in [21]. Also, traditional studies on the properties of
CCSs have focused on the description of the characteristics of the
codes in terms of shiftedness, BCD for numerics, BCD for
alphabetics, numerics in numeric sequence, signed numerics and the
matching of collating sequence with the bit sequence [25]. For
instance, the EBCDIC is not shifted and its alphabetics are not in
contiguous sequence. However, it posseses all the other properties
mentioned above . In some of his earlier papers [e.g 29, 31], the
author applied the idea of equivalence relation to ordinary
differential equations of the form $x^{/} = g(x) =
\sum_{i=1}^{n}a_{i}x^{n-i}, a_{i} \in \Re$ and constructed codes
using the quaternary system $\{Attractor(A), Repellor(R), Positive
Shunt(P) , Negative Shunt(N)\}$ where A, R, P and N are the
possible phase portraits for (a linear) equation on the line. The
blocklength of a  code constructed this way  is equal to the
degree n of the polynomial. In [6], a geometric technique for
constructing codes using the $black/white$ lift of a cap was
presented. This is accomplished by partitioning a set of points
which have no three of its points collinear. In the present paper
however, a new approach to the study of CCSs (and binary uniform
digital codes in general) is discussed and pertinent theorems on
the approach presented. The approach, called 'code presentation',
uses the concept of a partition and an imitation of the idea
behind 'group presentation'. This enables a CCS to be described in
a simple form in terms of a 'zoned set' and a 'decinumer set' just
as a group is described in terms of a 'set of generators' and a
'set of relators'[7, 12 , 14 , 18 , 22, 24 , 26, 27, 33]. In
particular, it is shown in [27] that a generalized free product of
two finitely presented groups acting on trees(a non-primitive
computer data structure) is finitely presented iff the amalgamated
subgroup is finitely generated. The general applicability of the
technique of code presentation to codes had been discussed in [30]
. As is well-known, the idea behind (group) presentation, in
itself, has been applied to a number of areas in mathematics and
computer science including geometry, topology, $C^{*}$-algebra,
knot theory, automorphic functions [26], computable algebra [15]
and other areas. Presentation in its ordinary meaning refers to
the depicting or writing of an algebraic structure in a simple
form. In particular, in the Theory of Computation , the
presentation of a function refers to a definition which gives an
effective method for computing the function [5]. In Formal
Language Theory, the concept of 'generation'(or derivation) is
associated with the phenomenon in which a language may be
generated by a phrase structure grammar [16]. Although the term
'Source Code Presentation' exists in Software Engineering, it is
used  in a different setting to describe the readability of the
source code of a program [35]. The beauty in group presentation ,
apart from its simplicity , is that it assists in deriving
information about an algebraic structure from its presentation.

\vspace{6mm} \noindent {\bf 2. $\,\,$  Group Presentation.}

\vspace{2mm}

Code presentation may be indirectly viewed as another area of
application of group theory. Generally, group presentation per se
is unsuitable for presenting CCSs since not all binary uniform
digital codes $(C, +)$ are groups where C is a typical code and
$'+'$ the binary operation defined on it. That is, the four basic
axioms of a group(namely closure, existence of an identity,
existence of an inverse and associativity) are not always
satisfied in a code . For instance, the simple code $\{00000,
01100, 00110, 11000, 11001, 11011\}$ of order six and blocklength
five is not a group because it is not closed. The methodology of
code presentation used in this paper is based on the premise  that
a typical modern day digital computer has an architecture in which
the bit patterns of a memory location in the main memory are
addressed in bytes i.e in multiples of bits.
    The following are well-known results on group presentation [8, 26] :
\vspace{2mm}
\noindent

{\bf Theorem 2.1.}

 (i) Every group has at least one set of generators, namely itself
 .\\
 (ii) Every group necessarily has a presentation. In particular, every
 finite group has a finite presentation. \\(iii) A subgroup of a finitely
presented group need not be finitely presented . \\(iv) A group
can have many presentations . \\(v) Given an arbitrary set of
symbols and an arbitrarily prescribed set of words in these
symbols, there exists a unique group, up to isomorphism, having
the symbols as generators and the set of prescribed words as
defining relators. (This result always allows new groups to be
constructed).\\

Two examples of group presentation satisfying the above are:\\
(i)Symmetric group of degree n$(S_{n})$ [22]\\
$$ S_{n} \approx  \{ \sigma_{1}\sigma_{2}...\sigma_{n-1} ; I_{n}\bigcup
B_{n} \}$$ where $I_{n}$  is the set of relations $\sigma_{i}^{2}
= 1$ for $i \in \{1,2,...,n-1\}$ and $B_{n}$ is the set of
relation $\sigma_{i}\sigma_{i+1}\sigma_{i} =
\sigma_{i+1}\sigma_{i}\sigma_{i+1}$ if $i \in\{1,2,...,n-1\}$ and
$\sigma_{i}\sigma_{j} = \sigma_{j}\sigma_{i}$ if $i , j \in
\{1,2,...,n-1 \}$ and $\vert {i - j} \vert \geq 2$. Under the
above isomorphism ,$\sigma_{i}$ is taken to be the transposition
$( i, i + 1)$. In particular, the presentation of the symmetric
group on three letters( $S_{3}$ ) is [3]\\ $$S_{3}\approx \{a , b
; a^{2} , b^{3}, a^{-1} bab^{-2}\}$$\\ (ii) The Mathieu group $M_{11}$ [12]\\
(a)$$ \{a,b,c,d,e ; aa, bb, cc, dd, ee, bdbd, bebe, (ab)^{3},
(de)^{3} ,(bc)^{5} , acece , a(cd)^{3}\}$$\\
(b)$$ \{a,b,c,d ; aa, bb , cc , dd , (ab)^{3} , (bc)^{3} ,
(cd)^{3} , (abdbd)^{2} , (cbdbd)^{3}\}$$\\
(c)$$ \{a, b, c, d ; aa , bb , cc , dd , acac , adad , (ab)^{3} ,
(bc)^{3} , (cd)^{3} , (bd)^{5} , (abd)^{5} , (bcdbcdc)^{3} \}$$

\vspace{6mm}
 \noindent
{\bf 3. $\,\,$ Code Presentation Theorems.}
\vspace{2mm}

Let $w = w_{1}w_{2}$ be a juxtaposed word of a uniform digital
code C of order k and blocklength n such that $w_{1} =
a_{1}a_{2}...a_{s}$ and $w_{2} = a_{s+1} a_{s+2}...a_{n}$. Then
$w_{1}$ and $w_{2}$ are respectively called the zoned portion and
numeric portion of w. If n is even we let $s = n/2$ and if $n$ is
odd, we let $s = ( n+1)/2$ or $s = ( n-1 )/2$. The Type I
definition of zoned portion for odd $n$ is depicted by the case
when $s = ( n+1)/2$ while the case in which   $s = ( n-1)/2$
describes the Type II definition of zoned portion. A constant
zoned portion refers to a zoned portion which is the same for two
or more words. Let the ordering on the code be according to the
collating sequence. Then the ordered set of all the constant zoned
portions of C is called the zoned code (or zoned set). The numer
code (or numer set) of C is the ordered set of all the numeric
portions of the code. We now suppose E is a subset of C. Then E is
said to be an equizone of C if all words in E have a single
constant zoned portion. The degree of C refers to the number of
equizones in it. A decinumer of E refers to the decimal value of a
numeric portion of E while the ordered set of all the decinumers
of an equizone is called a decinumer set [30].

\vspace{2mm}
\noindent
{\bf Theorem 3.1.}

 Every coded character set C can be well-ordered.
\vspace{2mm}
 \noindent

 {\bf Proof.}

A CCS is a set. By the well-ordering principle [20], every set can
be well-ordered. For $c_{1}, c_{2} \in C$, define $c_{1} \leq
c_{2}$ iff the decimal value of $c_1$ is less than or equal to the
decimal value of $c_{2}$ based on the collating sequence of C. It
is easily seen that C is a chain and every subset of it contains a
first element. Therefore, a CCS can be well-ordered .

\vspace{2mm}
 \noindent
 {\bf Definition 3.2(Fundamental Definition of Code Presentation).}

 Let a uniform digital code C has a degree d and suppose $E_{i}$ is a typical
 equizone of decinumer set  $Q_{i}$ where $i < d$.  If $z_{i}$ is the constant
 zoned portion of $E_{i}$ and $x_{ig}$ the bit pattern of $g \in Q_{i}$, then the
code presentation of C is given by:
$$C\{P\} = \bigcup_{i=1}^{d}\{z_i x_{ig} \forall g \in Q_{i}\}\eqno(3.2)$$
\vspace{2mm}
 \noindent
 {\bf Definition 3.3.}

Let $w_{1} = a_{11}a_{12}...a_{1n}$ and $w_{2} =
a_{21}a_{22}...a_{2n}$ be two words of a coded character set and
$'*'$ a binary operation. Then $w_{1}
* w_{2}$  is defined as
$$
w_{1} *w_{2}  = (a_{11} * a_{21} ) (a_{12}  * a_{22} ) …(a_{1n}
* a_{2n} )
\eqno{(3.3a)}
$$.
In particular,  the word difference( - ) between $w_1$ and $w_2$ ,
denoted by $w_{1}-w_{2}$, is given by
$$
w_{1} - w_{2} = (a_{11} - a_{21} ) (a_{12} - a_{22} ) …(a_{1n} -
a_{2n} ) \eqno{(3.3b)}
$$
where
$$a_{ij}-a_{kj}=\left\{\begin{array}{ll}
 0& \mbox{if}\;\;a_{ij}= a_{kj}\\
 1&\mbox{otherwise}\end{array}\right.$$
 \vspace{2mm}
 \noindent
 {\bf Remark 3.4.}

Addition in computer arithmetic is normally defined by:
    $0 + 0 = 0$ ; $0 + 1 = 1$ ; $1 + 0 = 1$ and $1 + 1 = 0$
while multiplication is normally defined by :
    $0 . 0 = 0$ ; $0 .1 = 0$ ; $1 .0 = 0$ and $1 . 1 =
    1$[9]
\vspace{2mm}
 \noindent

 {\bf Proposition 3.5.}
    $$w_{1} - w_{2}  = w_{1} + w_{2}$$

\vspace{2mm}
 \noindent
 {\bf Proof.}

Obvious; it follows from the fact that
    $a_{ij} - a_{kj}$ = $a_{kj}  - a_{ij}$  = $a_{ij}  +  a_{kj}$
\vspace{2mm}
 \noindent
 {\bf Theorem 3.6.}

Let C be a uniform digital code and T the zoned code of C. Given
$a, b \in  C$ and $z_{a}, z_{b} \in T $, let $a\sim b$ iff  $z_{a}
- z_{b} = 0$ , where $z_{a}$ , $z_{b}$  are respectively the zoned
portions of a and b and $'-'$ is the word difference. Then $\sim$
defines an equivalence relation \\

 {\bf Proof.}

This follows from the fact that $\sim$ is reflexive, symmetric and
transitive.\\

    The distinct equivalence classes of  $\sim$ are the equizones.
By virtue of the Fundamental Theorem of Equivalence Relations, it
follows that the set of all these equivalence classes gives a
partition of the code C.\\
{\bf Definition 3.7.}

Let $C= \{ X_{1}, X_{2},..., X_{e}\}$ be an ordered code of order
e  where $X_i = x_{i1}x_{i2}...x_{in}$ , $x_{ij} \in \{0,1\}$, $j
=1,2,...,n$ is a word of blocklength n . Then the inverse of C,
denoted by $C^{-1}$, is an ordered code given by $C^{-1}= \{ X_{n}
, X_{n-1}..., X_{1} \}$. The inverse of $X_{i}$, denoted by
$X_{i-1}$, is given by $X_{i}^{-1} = x_{in}
^{-1}x_{in-1}^{-1}...x_{i1}^{-1}$ where

$$x_{ij}^{-1}= \left\{\begin{array}{ll}
      0   & \mbox{if}\;\;x_{ij}= 1\\
      1   & \mbox{otherwise}\end{array}\right.$$

 {\bf Theorem 3.8.}

(i) Every coded character set(CCS) necessarily has a
presentation.\\ (ii)The order of the zoned set of a coded
character set is finite. \\(iii)Every subset of a coded character
set has a finite presentation.\\ (iv)The presentation of a coded
character set is not necessarily unique.\\ (v) A coded character
set can have at most two presentations.\\
 \vspace{2mm}
 \noindent

 {\bf Proof.}

(i) Since every CCS can be written in terms of a zoned portion and
a numeric portion , then the result follows. (ii)    A CCS is
finite . Therefore it has a finite number of zoned portions. (iii)
This follows from the fact that every subset of a CCS is finite.
(iv)    When a CCS has an odd blocklength , then its zoned
portion, by definition, can be described in two ways. It then
follows that  the CCS has two distinct presentations. (v) The two
distinct presentations in the proof of (iv) are the only possible
presentations. They are the maximum possible presentations.

\vspace{6mm}
\noindent

{\bf 4. $\,\,$ Example.}
 \vspace{2mm}

The  EBCDIC  is one of the two most popular CCSs in digital
computer systems [2]. It has 16  equizones  namely   $E_{0}$ ,
$E_{1} , ..., E_{9}$ , $E_{A}$ , $E_{B} ,...E_{F}$ .  The zoned
set of the code is the 4-bit hexad set  $H^{4}_{16} = \{0000,
0001, 0010, 0011,...,\\1110, 1111\}$ . Table 1 gives the decinumer
set of each equizone of the EBCDIC code where $'-'$ is the set
difference.  In the table, $l_{i}$  is the decinumer set which
corresponds to equizone $E_{i}$ while $T_{4}$ represents  the
order-preserving decinumer set of $H^{4}_{16}$ i.e $T_{4} = \{
0,1,2,..., 14, 15\}$. Both equizones $E_{1}$ and $E_{2}$ have the
highest  number of characters (i.e  13 characters each ) while
equizone $E_{B}$  with no character,  has the least number of
characters. \vspace{6mm} \noindent

{\bf 5. $\,\,$ Discussion and Conclusion.}

This paper has related the idea behind group presentation to coded
character sets(CCSs) in digital computer architecture . Theorems,
which are analogies of some of the well-known results in
combinatorial group theory,  are then presented. Just as group
presentation, code presentation enables CCSs to be written in a
simple form and also assists in deriving information about their
structure.  The results in the paper arise from the fact that
group presentation theory (or combinatorial group theory), is
generally unsuitable for presenting CCSs since not all CCSs are
groups. Apart from this, the bits of CCSs have some particular
pattern of representation in the classical computer hardware [23,
25].
\newpage
\noindent {\bf References.} \begin{description} \item[[1]] S.
Al-Bassam , Another Method for Constructing $t-EC{/}AUED$ Codes ,
IEEE Trans. Computers 49, No 9(2000) $964-966$.
 \item[[2]] B. R.Bannister , D. G. Whitehead , Fundamentals of Modern Digital
Systems , Macmillan Education Ltd, Basingstoke, Hampshire, 1987.
\item[[3]] B. Baumslag , B. Chandler , Schaum's Ouline of Theory
and Problems of Group Theory, McGrawHill Book Company, New York,
1968 \item[[4]] A.T. Berztiss, Data Structures:Theory and
Practice, Academic Press Inc., New York, 1975 \item[[5]] W. S.
Brainerd , L. Landweber , Theory of Computation , John Wiley \&
Sons Inc., 1974. \item[[6]] A. A Bruen , D. Wehlau ,  New Codes
from Old: A New Geometric
 Construction , J. Combin. Theory Ser A 94(2001) $196-202$.

\item[[7]] C. M. Campbell , I. Miyamoto , E.F.Robertson , P.D.
Williams ,
 The Efficiency of PSL$(2,p)^{3}$ and Other Direct Products of Groups ,
 Glasgow Math. J. 39(1997) $259-268$.

\item[[8]] C. O. Christenson , W. L. Voxman , Aspects of
Topology(Pure and Applied Mathematics - A Series of Monographs and
Textbooks), Marcel Dekker Inc., New York, 1977.

\item[[9]] A. Clements, The Principles of Computer Hardware
(Second
 Edition), Oxford University Press, 1991

\item[[10]] B. K. Dey, B. S. Rejan ,Codes Closed Under Arbitrary
Abelian Group
 of Permutations , SIAM J. Discrete Math. 18, No 1(2004) $1-18$.

\item[[11]] P. P. Dey ,Exploring Invariant Linear Codes Through
Generators and Centralizers , J. Discrete Math. Sci. \& Crypt  7,
No. 2(2004) $167-178$. \item[[12]] D. Z. Dokovic,  Presentations
of Some Finite Simple Groups , J. Austr. Math. Soc., Ser A
45(1988) $143-168$. \item[[13]] A. C. Downton , Computers and
Microprocessors: Components and Systems, Third edition, Chapman
and Hall, London , 1992. \item[[14]]
 V. Gebhardt , Constructing a Short Defining Set of Relations for a
Finite Group , J. Algebra 233(2000) 526-542. \item[[11]] S.
S.Goncharov,  S. Lempp, R. Solomon , The Computable Dimension of
Ordered Abelian Groups , Adv. Math. 175(2003) $102-143$.
\item[[16]] M. A. Harrison , Introduction to Formal Language
Theory , Addison-Wesley Publishing Company Inc., Philippines,
1978.

\item[[17]]  I. Herstein , Topics in Algebra , John Wiley \& Sons,
New York , 1975.

\item[[18]] S. Holub , Binary Equality Sets Are Generated by Two
Words , J. Algebra 259, No. 1(2003) $1-42$.

 \item[[19]] Y. Jang , J. Park, On a MacWilliams Type Identity and a Perfectness for
 a Binary Linear $(n,n-1,j)-$Poset Code , Discrete Mathematics 265(2003) $85-104$.

\item[[20]]  A. O. Kuku ,  Abstract Algebra, Ibadan University
Press ,1980.

 \item[[21]] C. Laih , C. Yang , On The Analysis and Design of
Group Theoretical $t-syEC{/}AUED$ Codes , IEEE Trans. Computers
45(1996) $103-108$.

\item[[22]] T. G. Lavers , Presentation of General Products of
Monoids , Journal of Algebra, 204(1998) $733-741$

 \item[[23]] S. Lipschutz , A.T. Poe, Schaum's Outline of Theory and Problems
of Programming With FORTRAN(Appendix B), McGrawHill Book Co,
Singapore, 1982.

\item[[24]] A. Lucchini , On The Number of Generators of Finite
Images of Free Products of Finite Groups , J. Algebra 245(2001)
$552-561$.

\item[[25]]  C. E. Mackenzie, Coded Character Sets, History and
Development(The Systems Programming Series), Addison-Wesley
Publishing Company, Philippines , 1980.

\item[[26]] W. Magnus, A.  Karrass, D. Solitar , Combinatorial
Group Theory(Presentations of Groups in Terms of Generators and
Relations), Dover Publications Inc, New York, 1976.

\item[[27]] R. M. S. Mahmood , On The Generators of the Edge
Stabilizers of Groups Acting on Trees , The Arabian Journal for
Science and Engineering 22, No. 2a(1997) $207-212$.

\item[[28]] D. Oluwade, Applications of 2-Code Error Detection
Techniques , Proceedings of the 14th National Conference of COAN
9(1998) $245-251$

\item[[29]] D. Oluwade , On the Qualitative Classes of $x^{/} =
ax^{3} + bx^{2} + cx + d$ , Proceedings of the National
Mathematical Centre, Abuja, Nigeria 1, No. 1(2000), $75-82$.

\item[[30]] D. Oluwade , Design and Analysis of Computer-Coded
Character Sets, Ph.D. thesis, University of Ibadan, November 2004.

\item[[31]] D. Oluwade , Modelling Fractal Patterns Via the
Qualitative Equivalence of a Nonlinear ODE , Nonlinear Analysis
63, No. $5-7(2005) e2409 - e2414$.

\item[[3]] D. Oluwade , Relating the Technique of Code
Presentation to the $K-Map$ Method in the Design of Code
Converters , Proceedings of the 3rd International Conference on
Computing, Communication and Control Technologies, Austin, Texas,
2005. \item[[33]] N. Ruskuc ,Presentations for Subgroups of
Monoids , J. Algebra 220(1999) $365-380$.

\item[[34]]  A. S. Tanenbaum ,Computer Networks , Prentice-Hall
Inc,
 Englewood Cliffs, New Jersey , 1989.

\item[[35]] $ www.doc-o-matic.com/doc/frames.html?frmname=topic$.
Retrieved in 2003.
\newpage
\begin{center}
TABLE 1: THE DECINUMER SET OF THE EBCDIC\\
\ \\
\begin{tabular}{|l|l|}\hline
 i &  $l_{i}$\\ \hline
 0 &  $T_{4}$ - $\{1,2,3,8,12\}$\\
 1 &  $T_{4}$ - $\{0,1,2\}$\\
 2 &  $T_{4}$ - $\{3,8,9\}$\\
 3 &  $T_{4}$ - $\{0,1,2,3\}$\\
 4 &  $T_{4}$ - $l_{8}$\\
 5 &  $l_{4}$\\
 6 &  $l_{4}\bigcup\{1\}$\\
 7 &  $l_{4}$- $\{0\}$\\
 8 &  $\{1,2,...,9\}$\\
 9 &  $l_{8}$\\
 A &  $l_{8}$- $\{1\}$\\
 B &  $\phi$\\
 C &  $l_{8}$\\
 D &  $l_{8}$\\
 E &  $l_{A}$\\
 F &  $l_{8}\bigcup \{0,15\}$\\\hline

\end{tabular}\\[2pt]
\end{center}

\newpage
LIST OF SYMBOLS\\
\begin{center}
\begin{tabular}{|l|l|}\hline
&\ \\
SYMBOL            &             MEANING\\ \hline
 $\in$            &             is an element of\\
 $\forall$          &             for all  \\
 $\leq$             &             less than or equal to\\
 $\bigcup$          &             union of some sets\\
 $H_{n}^{16}$     &             n-bit hexad set\\
 $\{\}$             &             set of elements\\
 $A-B$              &             the difference of two sets A and
                               B(set difference)\\
 $w_{1}- w_{2}$   &             word difference\\
 $T_4$            &             order-preserving equidext of $H_{4}^{16}$\\
 $\emptyset$        &             empty set\\\hline
 \end{tabular}\\[2pt]
 \end{center}

\end{description}
\end{document}